\documentclass[
twocolumn,
aps,
prl,
longbibliography
]{revtex4-1}

\usepackage{xcolor}

\usepackage{graphicx}
\usepackage{dcolumn}
\usepackage{bm}
\usepackage[english]{babel}
\usepackage{amsmath}
\usepackage{amssymb}

\begin{document}
\newcommand{\vb}[1]{\boldsymbol{#1}}
\newcommand{\expval}[1]{\langle #1\rangle}
\newcommand{\mat}[1]{\boldsymbol{\mathrm{#1}}}
\newcommand{\N}{N}
\newcommand{\K}{K}
\newcommand{\n}{n}
\newcommand{\T}{\mathcal{T}}
\newcommand{\occ}[2]{\n_{#1}(#2)}
\newcommand{\conf}{\Gamma}
\newcommand{\Size}{L}
\newcommand{\Q}{\mathcal{Q}}
\newcommand{\Xtype}[2]{#1_{#2}}
\newcommand{\Xindex}[2]{#1^{(#2)}}
\newcommand{\QMTUR}{\mathcal{Q}_{\vb{J}}}
\newcommand{\C}{\mathcal{C}}
\newcommand{\Cmat}{\mat{\C}}
\newcommand{\set}[1]{\{#1\}}
\newcommand{\sigest}{\sigma_\mathrm{est}}
\newcommand{\sigestMTUR}{\sigma_\mathrm{est}^{\vb{J}}}
\newcommand{\todo}[1]{ {\textcolor{red}{#1}} }
\newcommand{\R}[1]{R_{#1}}
\newcommand{\NN}[1]{\mathcal{N}\left(#1\right)}

\preprint{APS/123-QED}

\title{Thermodynamic Uncertainty Relation in Interacting Many-Body Systems\\~\\}

\author{Timur Koyuk and Udo Seifert\\~}
\affiliation{
{II.} Institut f\"ur Theoretische Physik, Universit\"at Stuttgart,
  70550 Stuttgart, Germany
}
\date{\today}

\begin{abstract}
The thermodynamic uncertainty relation (TUR) has been well studied for systems
with few degrees of freedom. While, in principle, the TUR holds for more
complex systems with many interacting degrees of freedom as well, little is
known so far about its behavior in such systems.  We analyze the TUR in the
thermodynamic limit for mixtures of driven particles with short-range
interactions. Our main result is an explicit expression for the optimal
estimate of the total entropy production in terms of single-particle currents
and correlations between two-particle currents. Quantitative results for
various versions of a driven lattice gas demonstrate the practical
implementation of this approach.
\end{abstract}

\maketitle
\textsl{Introduction.}
Fluctuating currents and their correlations are a characteristic signature of
stationary non-equilibrium systems. Exact results like the fluctuation
theorem~\cite{evan93,gall95,kurc98,lebo99,andr07c} and the Harada-Sasa
relation~\cite{hara05} represent prominent, universal predictions that relate
currents and their correlations to the arguably most central quantity for such
systems -- the rate of entropy production. More recently, the thermodynamic
uncertainty relation (TUR)~\cite{bara15,ging16,horo20} has revealed an unexpected
constraint on the precision of any current in terms of the total entropy production
rate.  Being a trade-off relation between precision and thermodynamic cost, in
the sense that a high precision requires a large amount of entropy production,
the TUR provides valuable insights into small mesoscopic non-equilibrium
systems. It has opened a variety of promising applications like for molecular
motors~\cite{piet16b}, heat engines~\cite{shir16,piet17a,holu18,ekeh20},
optimal design principles for self-assembly~\cite{nguy15} or constraints on time
windows in anomalously diffusing systems~\cite{hart21}. From the perspective
of thermodynamic inference, being a simple tool for estimating entropy
production by measuring experimentally accessible currents and their
fluctuations without knowing interaction potentials or driving forces, the TUR
has been established as an indispensable addition to more sophisticated inference
methods~\cite{brue20,fris20,skin21,skin21a}.

To explore these two key properties of the TUR in more complex situations,
subsequent work has focused on extending its range of applicability to a
variety of systems including the observation of steady states in finite
times~\cite{piet17,horo17}, underdamped
dynamics~\cite{dech18,chun19,fisc19,lee19,vanv19,lee21,kwon21,piet21},
stochastic field theories~\cite{nigg22}, observables that are even under
time-reversal~\cite{maes17,nard17a,terl18}, first-passage
times~\cite{ging17,garr17}, relaxation processes~\cite{dech18a,liu19,wolp20},
periodically~\cite{bara18b,koyu19a} and arbitrary time-dependently driven
systems~\cite{vanv20,koyu20}. Several of these generalizations have
been (re-)derived by using virtual perturbations or information
theoretic bounds~\cite{dech17,dech18a}.  Last but not least, various studies
have worked on generalizations of the TUR to open quantum
systems~\cite{maci18,agar18,ptas18,bran18,carr19,guar19,caro19,pal20,frie20,hase21,mill21}.

When dealing with generalization and refinements of the TUR, a crucial
question is how sharp the corresponding bounds typically are. Early analyses
showed that the TUR can be saturated in the linear response regime due to
Gaussian fluctuations~\cite{bara15,ging16,piet15,ging16a}. More recent studies
have revealed that for the same reason it can become tight in the short-time
limit~\cite{dech20,mani20}. The same situation has been observed for the
time-dependent TUR in the fast-driving
limit~\cite{koyu20,koyu21}. In all these cases, only the current of total
entropy production, or a current proportional to it, leads to an equality in
the TUR. Further works have focused on finding the optimal observable(s)
leading to the tightest possible
bound~\cite{pole16,ging16a,busi19,li19,fala20,mani20,shir21}.  More
specifically, using a sum of two observables and, thus, using correlations
between them, can yield a sharper bound~\cite{dech21}.

Most of the specific studies so far have treated single-particle systems or
systems with a few degrees of freedom on a mesoscopic scale.  As a crucial
refinement of the TUR, the multidimensional thermodynamic uncertainty relation
(MTUR)~\cite{dech18b} should become useful when dealing with multiple currents
and their correlations.  While this refinement provides, in principle, the
possibility to analyze systems with many interacting degrees of freedom, a
systematic study of the thermodynamic limit is still missing.

In this Letter, we analyze the TUR in the thermodynamic limit and derive the
optimal estimate of entropy production using the MTUR. Our results hold for
any driven many-particle system obeying a Markovian dynamics on a discrete set
of states or overdamped Langevin equations.  We will illustrate our
theoretical predictions with various versions of a driven lattice gas.

\textsl{TUR in many-particle systems.}
The thermodynamic uncertainty relation has been proven under quite general
conditions for both continuous-time Markov-processes and systems obeying
coupled overdamped Langevin equations.  It is valid for any current and
reads~\cite{bara15,ging16,horo20}
\begin{equation}
  \label{eq:intro_TUR}
  \sigest^J \equiv J^2/D_J \le \sigma,
\end{equation}
where $J$ is the mean current, $D_J \equiv \T \mathrm{Var}[J]/2$ is its
diffusion coefficient, and $\T$ denotes the observation time~\footnote{See
Supplemental Material at [SI] for definitions of currents, diffusion
coefficients, covariances, and the full derivations of the main results.}.
The precision $J^2/D_J$ bounds the total entropy production rate $\sigma$ and,
hence, yields an operationally accessible estimate $\sigest^J$ for it. To
analyze the sharpness of the TUR, we define the quality factor as
$\Q_J\equiv\sigest^J/\sigma > 0$, which is $1$ if the TUR is saturated. Since
the TUR~\eqref{eq:intro_TUR} holds for any current in the system, we can use
an arbitrary linear combination of currents to build the estimate $\sigest^J$.

To study the TUR of interacting many-particle systems, we use a refinement of
the TUR -- the so-called MTUR introduced in Ref.~\cite{dech18b}. We consider a
system that consists of $\N$ driven interacting particles leading to $\N$
linearly independent particle currents $\set{\Xindex{J}{i}}$.  The MTUR can be
applied by inserting the optimal linear combination of these currents
into~\eqref{eq:intro_TUR}. Within this class of currents, it thus yields the
sharpest lower bound on the entropy production, which is given by
\begin{equation}
  \label{eq:MTUR:MTUR} \sigestMTUR \equiv \vb{J}^{T}\Cmat^{-1}\vb{J} \le
\sigma.
\end{equation}
The estimator $\sigestMTUR$ involves the
vector of particle currents $\vb{J}\equiv\left(\Xindex{J}{1},...,\Xindex{J}{\N}\right)$ and the inverse of the
symmetric correlation matrix $\Cmat$ with elements
\begin{equation}
  \label{eq:MTUR:correlation_matrix}
  \C_{ij} \equiv \Xindex{D}{i} \delta_{i,j} + \Xindex{C}{ij}(1-\delta_{i,j}).
\end{equation}
The diagonal element $\Xindex{D}{i}\equiv\T\mathrm{Var}[\Xindex{J}{i}]/2$ is
the diffusion coefficient of the current $\Xindex{J}{i}$ of the $i$th particle
and the off-diagonal elements
$\Xindex{C}{ij}\equiv\T\mathrm{Cov}[\Xindex{J}{i},\Xindex{J}{j}]/2$ are the
scaled covariances between the currents $\Xindex{J}{i}$ and
$\Xindex{J}{j}$~\cite{Note1}.

We use the MTUR to obtain the optimal estimate for entropy production in
the thermodynamic limit for a system with different species of
particles. First, we consider a homogeneous system with only one
species driven by a thermodynamic force $f$. Here, all mean
values $\Xindex{J}{i} \equiv J$, diffusion coefficients $\Xindex{D}{i} \equiv
D$ and correlations $\Xindex{C}{ij} \equiv C$ are identical.
Since all particles are indistinguishable each current contributes with the
same weight to the optimal linear combination such that the MTUR reduces to
the ordinary TUR for the total particle current. Hence, the estimate is given
by~\cite{Note1}
\begin{equation}
  \label{eq:MTUR:estimate_one_species}
  \sigestMTUR = \frac{NJ^2}{D - C + N C},
\end{equation}
while the true entropy production reads $\sigma = \beta f N J$ with the
inverse temperature $\beta$ and $k_\mathrm{B}=1$.  In the thermodynamic limit
$N\to\infty$, the correlations generically decay like $C\approx \gamma/N$ with
amplitude $\gamma$ since the probability to find two labelled particles near each other
is proportional to the system size and, hence, at fixed density to the number of particles.
When taking the thermodynamic limit $N\to\infty$, the quality factor becomes
\begin{equation}
  \label{eq:MTUR:quality_factor_one_species_thermodynamic_limit}
  \QMTUR \equiv \frac{\sigestMTUR}{\sigma} = \frac{J^\infty}{\beta f (D^\infty + \gamma)},
\end{equation}
where $J^\infty$ and $D^\infty$ are the value of the particle current and its
diffusion coefficient in the thermodynamic limit, respectively.
Equation~\eqref{eq:MTUR:quality_factor_one_species_thermodynamic_limit} is our
first main result and shows
that the quality of the estimate depends solely on one- and two-particle
quantities. For a driven ideal gas ($\gamma=0$), the currents become uncorrelated and
the $\N$-particle system corresponds to a single-particle system. In contrast, for
strong correlations between the particle currents, i.e., large absolute
correlation amplitudes $\gamma$, the quality factor can differ strongly from the
single-particle case.

Next, we study a homogeneous system consisting of a mixture of $\Xtype{\N}{1}$
particles of species $1$ and $\Xtype{N}{2}$ particles of species $2$. The
first and second species are driven by forces $\Xtype{f}{1}$ and
$\Xtype{f}{2}$, respectively. Particles interact with a short-range
interaction, which may be different between the species. The
mean particle currents within a species are identical, i.e., $\Xindex{J}{i}=
\Xtype{J}{\alpha_i}$, where $\alpha_i\in\set{1,2}$ denotes the species of the
$i$th particle. Analogously, the diffusion coefficients
$\Xindex{D}{i}=\Xtype{D}{\alpha_i}$ and correlations
$\Xindex{C}{ij}=\Xtype{C}{\alpha_i \alpha_j}$ depend only on the particle
species.
Using
Eq.~\eqref{eq:MTUR:MTUR} we get the estimate~\cite{Note1}
\begin{equation}
  \label{eq:MTUR:estimate_two_species}
  \sigestMTUR = \frac{\Xtype{\eta}{2}\Xtype{\N}{1}\Xtype{J}{1}^2 + \Xtype{\eta}{1}\Xtype{N}{2}\Xtype{J}{2}^2 -
    2\Xtype{\N}{1}\Xtype{N}{2}\Xtype{J}{1}\Xtype{J}{2}\Xtype{C}{12}}{\Xtype{\eta}{1}\Xtype{\eta}{2} - \Xtype{\N}{1}\Xtype{N}{2} \Xtype{C}{12}^2}
\end{equation}
with $\Xtype{\eta}{\alpha} \equiv \Xtype{D}{\alpha} +
(\N_\alpha-1)\Xtype{C}{\alpha\alpha}$ and $\alpha\in\set{1,2}$.
The true entropy production is given by $\sigma = \beta
\Xtype{f}{1}\Xtype{\N}{1}\Xtype{J}{1} + \beta \Xtype{f}{2}\Xtype{N}{2}\Xtype{J}{2}$. When taking the
thermodynamic limit $\N = \Xtype{\N}{1} + \Xtype{N}{2} \to \infty$, we keep the densities
$\Xtype{\rho}{\alpha} \equiv \Xtype{N}{\alpha}/\N$ fixed. Analogously to the one-species case,
the correlations $\Xtype{C}{11}\approx \Xtype{\gamma}{1}/N$,
$\Xtype{C}{22}\approx\Xtype{\gamma}{2}/N$ and $\Xtype{C}{12}\approx\Xtype{\gamma}{12}/N$
decay proportional to the inverse system size, where $\Xtype{\gamma}{1}$,
$\Xtype{\gamma}{2}$ and $\Xtype{\gamma}{12}$ are the correlation
amplitudes. Thus, the quality factor in the thermodynamic limit reads
\begin{equation}
  \label{eq:MTUR:quality_factor_two_species_thermodynamic_limit}
  \QMTUR = \frac{\Xtype{\eta}{2}^\infty\Xtype{\rho}{1}(\Xtype{J}{1}^\infty)^2
    + \Xtype{\eta}{1}^\infty\Xtype{\rho}{2}(\Xtype{J}{2}^\infty)^2 -
    2\Xtype{J}{1}^\infty\Xtype{J}{2}^\infty\Xtype{\rho}{1}\Xtype{\rho}{2}\Xtype{\gamma}{12}}{\left[\Xtype{\eta}{1}^\infty\Xtype{\eta}{2}^\infty-\Xtype{\rho}{1}\Xtype{\rho}{2}\Xtype{\gamma}{12}^2\right]\beta\left(\Xtype{f}{1}\Xtype{\rho}{1}\Xtype{J}{1}^\infty
    + \Xtype{f}{2}\Xtype{\rho}{2}\Xtype{J}{2}^\infty\right)},
\end{equation}
where $\Xtype{\eta}{\alpha}^\infty \equiv \Xtype{D}{\alpha}^\infty +
\Xtype{\rho}{\alpha}\Xtype{\gamma}{\alpha}$ and $\Xtype{J}{1,2}^\infty$ and
$\Xtype{D}{1,2}^\infty$ denote the values of the currents and diffusion
coefficients in this limit, respectively. The quality factor in
Eq.~\eqref{eq:MTUR:quality_factor_two_species_thermodynamic_limit} is our
second main result. In contrast to the one species case, this quality factor
differs from the quality factor obtained by using as a current the total
power~\cite{Note1}. In both cases, the MTUR remains a useful tool to
infer entropy production, which, in particular, does not require the knowledge
of any thermodynamic forces as we will now illustrate for the driven lattice
gas.

\begin{figure}[tbp]
  \centering
  \includegraphics[width=0.5\textwidth]{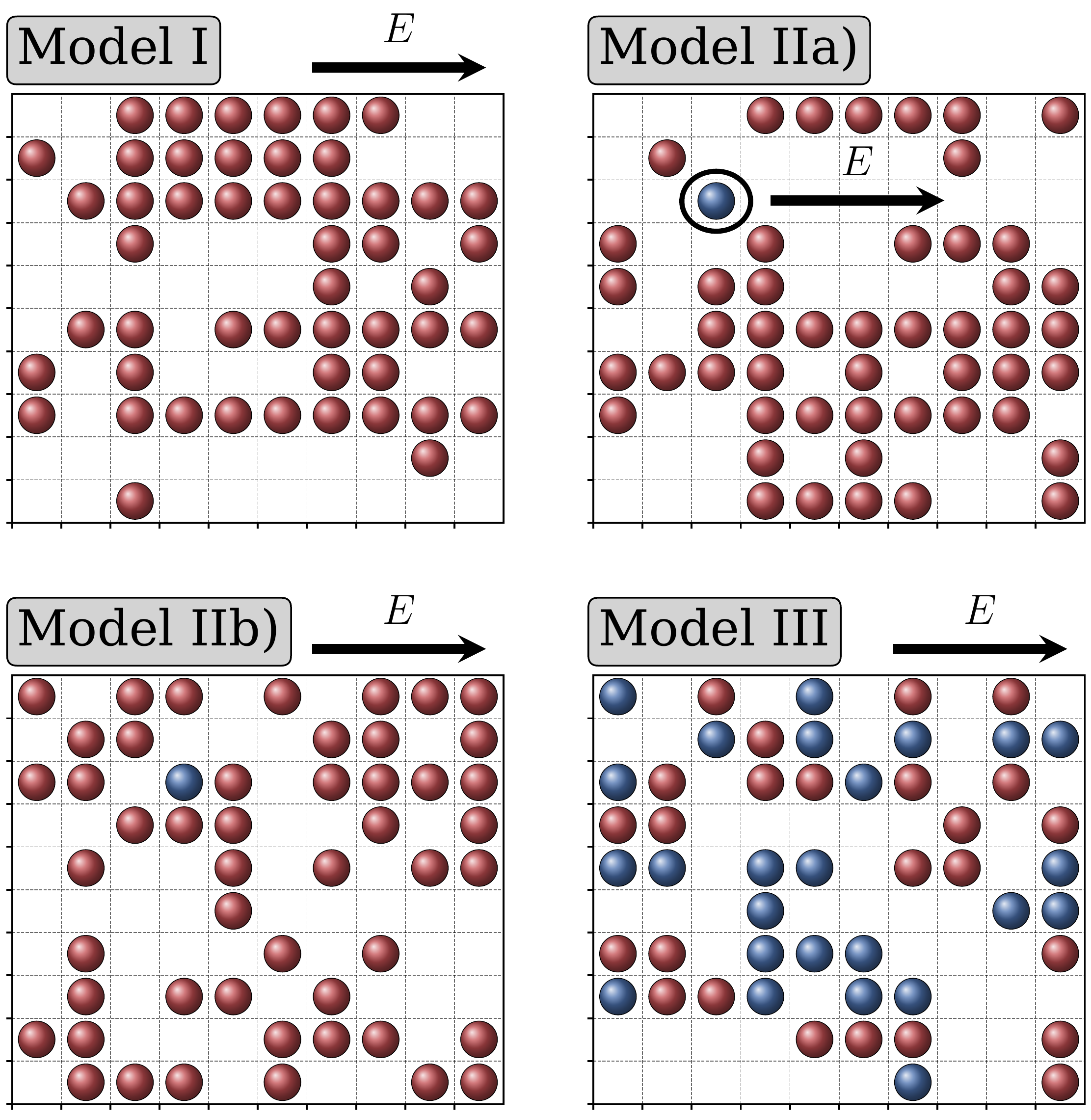}
  \caption{Three different models of the driven lattice gas I-III. In model
    IIa) there is only one particle (the blue one) driven by the
    electric field, whereas in model IIb) all particles are driven.}
\label{fig:DLG_models_illustration}
\end{figure}

\textsl{Driven lattice gas.}
We consider a driven lattice gas~\cite{katz84} in which $\N$ charged particles
occupy sites on a periodic $\left(\Size\times\Size\right)$--square lattice
subject to an exclusion interaction as shown in
Fig.~\ref{fig:DLG_models_illustration}.
The particles are driven by an electric
field applied in the $x$-direction. Moreover, each
particle interacts with its nearest neighbors either repulsively or
attractively. The occupation variable $\occ{i}{\vb{r}}\equiv
\delta_{\vb{r}_i,\vb{r}}$ at position $\vb{r} \equiv(x,y)$ is one,
if particle $i$ at $\vb{r}_i\equiv(x_i, y_i)$ occupies this site and is zero,
otherwise.  The configuration of the system is denoted by
$\conf \equiv \set{\occ{i}{\vb{r}}}$, which contains information about all
particle positions $\R{\conf}\equiv\set{\vb{r}_1,...,\vb{r}_\N}$. 

In the following, we consider a system consisting of two species of particles
with different charges $q_1$ and $q_2$. The
interaction energy of the total system is given by
\begin{equation}
  \label{eq:DLG:interaction_energy}
  E_\mathrm{int}(\conf) \equiv -\sum_{i>j} \Xtype{\K}{\alpha_i\alpha_j} \sum_{<\vb{r}\vb{r'}>} \occ{i}{\vb{r}}\occ{j}{\vb{r'}},
\end{equation}
where $\sum_{<\vb{r}\vb{r'}>}$ denotes a summation over all
nearest-neighbor-site pairs and
$\Xtype{\K}{\alpha_i\alpha_j}$ is the coupling constant of species
$\alpha_i$ and $\alpha_j$ with $\alpha_{i,j}\in\set{1,2}$. If
$\Xtype{\K}{\alpha_i\alpha_j} > 0$ the interaction is attractive, otherwise repulsive.
The probability $p(\conf,t)$ to find the system in
configuration
$\conf$ at time $t$ obeys the master equation
\begin{align}
  \label{eq:DLG:master_equation} \partial_t p(\conf, t) =
  \sum_{\substack{\vb{r}_i\in\R{\conf}, \\
      \vb{r'}_i\in\NN{\vb{r}_i}} } &\Big[p(\conf^{\vb{r}_i\vb{r'}_i},t)
  k(\vb{r'}_i, \vb{r}_i, \conf^{\vb{r}_i\vb{r'}_i})\nonumber \\  &- p(\conf,t)
k(\vb{r}_i, \vb{r'}_i, \conf)\Big],
\end{align}
where $\NN{\vb{r}_i}$ denotes a set of all unoccupied nearest neighbor
sites $\vb{r'}_i\equiv(x'_i, y'_i)$ of position $\vb{r}_i$ and
$\conf^{\vb{r}_i\vb{r'}_i}$ denotes a configuration identical to $\conf$
except that particle $i$ occupies $\vb{r'}_i$ instead of $\vb{r}_i$.
The transition rate for a
particle at $\vb{r}_i$ to move to an unoccupied nearest neighbor site
$\vb{r'}_i$ fulfills the local detailed
balance condition and is given by
\begin{equation}
  \label{eq:DLG:transition_rate}
  k(\vb{r}_i, \vb{r'}_i, \conf) \equiv
  \begin{cases}
k_0\exp\left(-\beta\kappa\Delta F\right), &\Delta F \ge 0\\
k_0\exp\left(\beta[1-\kappa]\Delta F\right), &\Delta F < 0
  \end{cases}
\end{equation}
with
\begin{equation}
  \label{eq:DLG:Delta_F} \Delta F \equiv
E_\mathrm{int}(\conf)-E_\mathrm{int}(\conf^{\vb{r}_i\vb{r'}_i}) +
(x_i-x_i') q_{\alpha_i} E.
\end{equation}
The rate amplitude $k_0$ determines the time scale for a transition,
$q_{\alpha_i}\in\{q_1, q_2\}$ denotes the charge of the moving particle, and
the parameter $\kappa$ determines the rate splitting.

\textsl{Quality factors.} 
We now analyze three paradigmatic models as depicted in
Fig.~\ref{fig:DLG_models_illustration}.
Model I consists of a single particle species with density $1/2$.  Model II
consists of $\Xtype{\N}{1}$ particles of species $1$ (red) and one single
particle of species $2$ (blue). Here, we distinguish two subclasses of models,
which we denote as IIa) and IIb): in model IIa) only the single particle of
species $2$ is charged, i.e., $q_1=0$ and $q_2\neq 0$, whereas in model IIb)
all particles are charged with, in general,
$\Xtype{q}{1}\neq\Xtype{q}{2}$. The number of particles of the first species
is chosen such that the density is $1/2$. Model III consists of two species
(red and blue particles) with different charges, interactions, and densities
$\Xtype{\rho}{1} = \Xtype{\rho}{2} = 1/4$.  For all models, we fix the
parameters $k_0=0.5$, $\beta=1.0$, $E=1.0$ and $\kappa=1.0$ and choose an
attractive interaction, i.e., $\Xtype{K}{11}, \Xtype{K}{22}, \Xtype{K}{12} >
0$. Moreover, we choose an observation time of $\T=1000.0$ to sample
trajectories by using the Gillespie
algorithm~\cite{gill77}. In the following, we analyze these systems for
different system sizes $\Size\times\Size$ and an overall density of $1/2$.

\begin{figure}[tbp]
  \centering
  \includegraphics[width=0.5\textwidth]{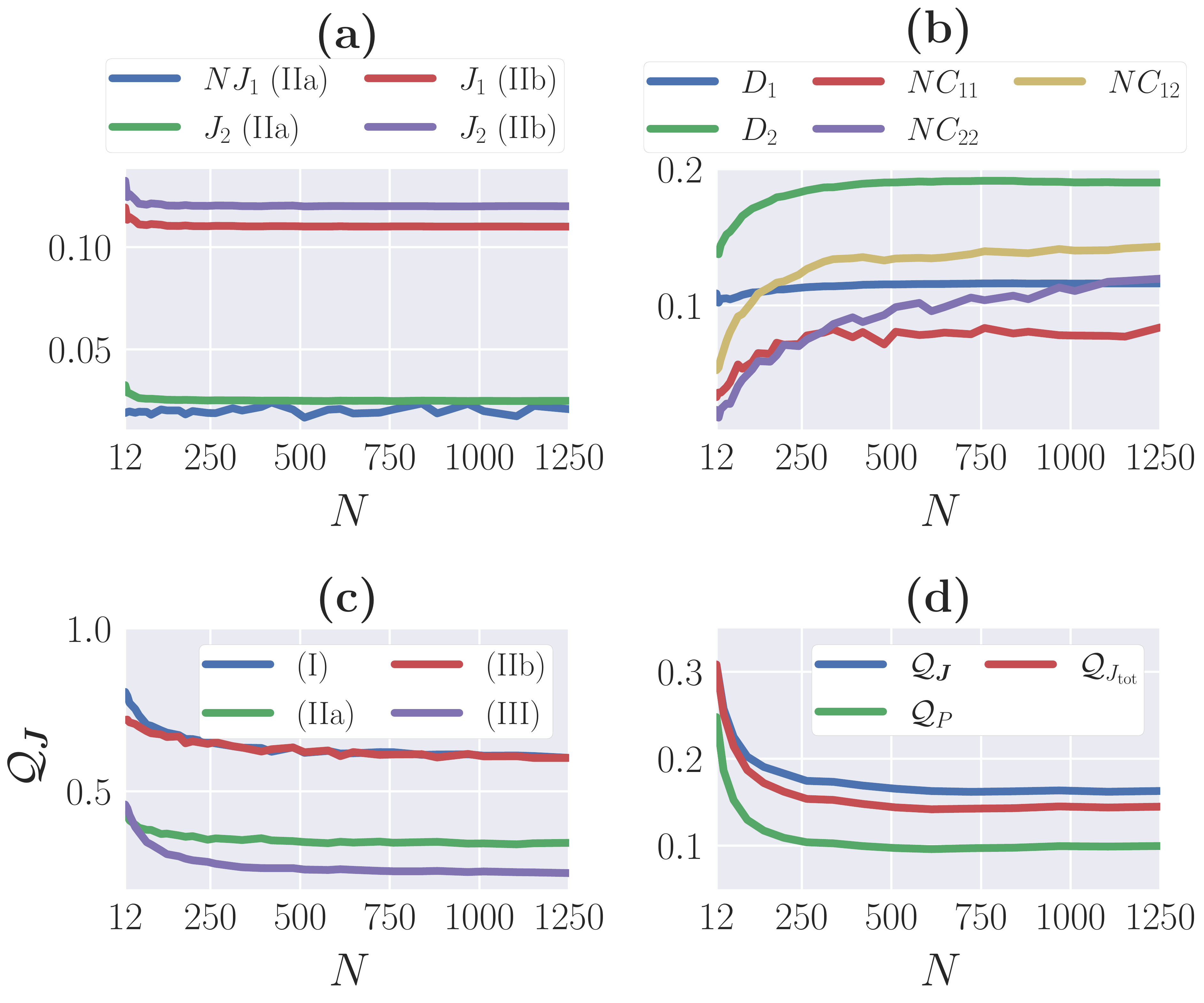}
  \caption{(a) Particle currents of models IIa) and IIb), (b) diffusion
coefficients and correlations of model III, (c) quality factors of all models
against $\N$ and (d) different quality factors $\QMTUR$, $\mathcal{Q}_\sigma$
and $\mathcal{Q}_{J_\mathrm{tot}}$ for model III. For model I, $q_1=1.0$ and
$\Xtype{\K}{11}=0.8$. For models IIa) and IIb), $q_1=0$, $q_2=1.0$,
$\Xtype{\K}{11}=0.8$ and $\Xtype{K}{12}=1.2$ and $q_1=1.0$, $q_2=2.5$,
$\Xtype{\K}{11}=0.8$ and $\Xtype{K}{12}=1.2$, respectively. For model III in
(c), $q_1=1.0$, $q_2=2.5$, $\Xtype{\K}{11}=0.8$, $\Xtype{K}{12}=1.2$ and
$\Xtype{K}{22}=0.4$, whereas in (d) $q_1=1.5$, $q_2=5.0$,
$\Xtype{\K}{11}=0.8$, $\Xtype{K}{12}=1.2$ and $\Xtype{K}{22}=0.4$.}
\label{fig:DLG_models_J_D_Q}
\end{figure}

Figure~\ref{fig:DLG_models_J_D_Q}(a) shows the single-particle currents
$J_1$ and $J_2$ of the two species for the models IIa) and IIb).
The single driven particle in model IIa) generates a particle current
$\Xtype{J}{1}$ of the $\Xtype{\N}{1}$ non-driven particles by pushing or
pulling them in the $x$-direction through the exclusion interaction and an
attractive short-range interaction.  Since this push- and pull-mechanism is a
local effect, the number of pushed or pulled particles saturates in the
thermodynamic limit, whereas the system size grows linearly in $\N$. As a
consequence, the current $\Xtype{J}{1}$ vanishes like $1/\N$, whereas the
current $\Xtype{J}{2}$ of the driven particle is finite as shown in
Fig.~\ref{fig:DLG_models_J_D_Q}(a). The optimal estimate of entropy production
can be obtained by setting $\Xtype{N}{2}=1$ and $\Xtype{C}{22}=0$ in
Eq.~\eqref{eq:MTUR:estimate_two_species}. The true entropy production is given
by $\sigma = \beta \Xtype{q}{2}E\Xtype{J}{2}$. Combined with the fact that
$\Xtype{J}{1}\sim 1/\N$ vanishes, these results imply that, in the
thermodynamic limit $\N\to\infty$, the quality factor becomes the quality
factor of a single-particle problem $\QMTUR = \Xtype{J}{2}^\infty/(\beta q_2
E\Xtype{D}{2}^\infty)$. The correlations between the particles do not
contribute to the quality factor in contrast to model I with one single
species
[cf. Eq.~\eqref{eq:MTUR:quality_factor_one_species_thermodynamic_limit}].

In model IIb), both particle currents $\Xtype{J}{1,2}$ are finite in the
thermodynamic limit as shown in Fig.~\ref{fig:DLG_models_J_D_Q}(a). The
quality factor can analogously be obtained by setting formally
$\Xtype{N}{2}=1$ and $\Xtype{C}{22}=0$ in
Eq.~\eqref{eq:MTUR:estimate_two_species} and using $\sigma = \beta
\Xtype{\N}{1} q_1 E \Xtype{J}{1} + \beta q_2 E \Xtype{J}{2}$. When taking the
thermodynamic limit $\N = \Xtype{N}{1} + 1\to\infty$, only quantities of the
first species contribute such that the quality factor reduces to the quality
factor of a single interacting species of particles, i.e.,
$\QMTUR=\Xtype{J}{1}/(\beta q_1
E[\Xtype{D}{1}+\Xtype{\gamma}{11}])$. Therefore, the quality factors of the
one-species system (a) and of system IIb) converge to the same value for
$\N\to\infty$ as illustrated in Fig.~\ref{fig:DLG_models_J_D_Q}(c), which
shows the quality factors for the different models I-III.

In model III, the diffusion coefficients $\Xtype{D}{1}$ and $\Xtype{D}{2}$ of
the two species converge to finite values as shown in
Fig.~\ref{fig:DLG_models_J_D_Q}(b). The correlations $\Xtype{C}{11}$,
$\Xtype{C}{22}$ and $\Xtype{C}{12}$ decay like $1/\N$ for the same reason as
above. These correlations multiplied with $\N$ are shown in
Fig.~\ref{fig:DLG_models_J_D_Q}(b).  Furthermore, for a small number of
particles $\N\lesssim 12$, model III becomes similar to model IIa): in both
models, species $1$ is either not driven or more weakly driven in contrast to
species $2$, which is strongly driven. This explains why the quality factors
of both models in Fig.~\ref{fig:DLG_models_J_D_Q}(c) approach each other for
small $\N$. However, for large $\N$ model IIa) is effectively a
single-particle problem and differs substantially from model III, in which
many driven particles interact. Thus, the quality factor reaches the larger
value $\QMTUR\simeq 0.38$ for model IIb), whereas it reaches $\QMTUR\simeq
0.30$ for model III. Most importantly, even though all quality factors shown
in Fig.~\ref{fig:DLG_models_J_D_Q}(c) decrease monotonically in $\N$, they
approach a finite value of order $1$ in the thermodynamic limit. In this
limit, model III has the smallest quality factor since the particles are
driven more strongly due to larger charges. Stronger driving leads to a smaller
quality factor since the particle currents and their fluctuations saturate for
large driving due to the exclusion interaction~\cite{katz84} while the entropy
production increases.

\textsl{Inference of entropy production.}
We finally compare different estimates of the entropy production for the most
interesting model III. The optimal quality factor $\QMTUR$ obtained from the
MTUR~\eqref{eq:MTUR:MTUR}, the quality factor using the total power as a
current $\mathcal{Q}_P$, and the quality factor $\mathcal{Q}_{J_\mathrm{tot}}$
of the total particle current $J_\mathrm{tot}\equiv \Xtype{\N}{1}\Xtype{J}{1}
+ \Xtype{\N}{1}\Xtype{J}{2}$ are plotted against $\N$ in
Fig.~\ref{fig:DLG_models_J_D_Q}(d).  As expected, the quality factor $\QMTUR$
beats the other two. The quality factor based on the power is even smaller
than $\mathcal{Q}_{J_\mathrm{tot}}$ and reaches a finite value of order $1$.
This is quite remarkable since it shows that the additional knowledge of
thermodynamic forces entering the power does not yield a better estimate. A
situation related to ours has been discussed in Ref.~\cite{busi19}, where the
authors have optimized a state-dependent increment for a current and found
that the best estimate does not coincide with the total entropy production. In
contrast to their approach, we use constant increments and build the optimal
linear combination of currents via the MTUR.

\textsl{Conclusion.}
In this Letter, we have analyzed the thermodynamic uncertainty relation for
interacting many-particle systems in the thermodynamic limit. We have
calculated the quality factor using the MTUR for a homogeneous system
consisting of a single species of particles and for a mixture of two species.
As we have shown the TUR remains a useful tool for inferring entropy
production since, crucially, the quality factors approach a finite order of
$1$ in the thermodynamic limit. From an operational perspective, it is neither
necessary to know the driving fields nor the interactions between the
particles in order to deduce the optimal estimate for entropy production. It suffices to measure the currents
and correlations between two different ones. Even though these correlations
vanish asymptotically, they are an essential ingredient to the TUR in systems
with many degrees of freedom.

With these results, we have laid the foundation for future studies of the TUR in
more complex systems, e.g., in systems with different phases
or at a phase transition. We stress that the analytical
results, Eqs.~\eqref{eq:MTUR:estimate_one_species}--\eqref{eq:MTUR:quality_factor_two_species_thermodynamic_limit},
apply to continuous overdamped Langevin systems as well. In this context, our
model IIa) corresponds to a driven Brownian particle embedded in a
colloidal suspension. Analyzing the TUR for different interaction potentials
or for systems with more than two species is an important next step to explore
macroscopic effects of the TUR. Since our tools rely on a widely applicable
mathematical framework, our results should open the way for future research to
study the thermodynamic limit of generalizations of the TUR, e.g., for
time-dependently driven systems or for open quantum systems.

\bibliography{../Bibliography/refs.bib}
\end{document}


\newcommand{\vb}[1]{\boldsymbol{#1}}
\newcommand{\dd}[1]{\mathrm{d}{#1}\,}
\newcommand{\expval}[1]{\langle #1\rangle}
\newcommand{\mat}[1]{\boldsymbol{\mathrm{#1}}}
\newcommand{\N}{N}
\newcommand{\K}{K}
\newcommand{\n}{n}
\newcommand{\T}{\mathcal{T}}
\newcommand{\occ}[2]{\n_{#1}(#2)}
\newcommand{\conf}{\Gamma}
\newcommand{\Size}{L}
\newcommand{\Q}{\mathcal{Q}}
\newcommand{\Xtype}[2]{#1_{#2}}
\newcommand{\Xindex}[2]{#1^{(#2)}}
\newcommand{\QMTUR}{\mathcal{Q}_{\vb{J}}}
\newcommand{\C}{\mathcal{C}}
\newcommand{\Cmat}{\mat{\C}}
\newcommand{\set}[1]{\{#1\}}
\newcommand{\sigest}{\sigma_\mathrm{est}}
\newcommand{\sigestMTUR}{\sigma_\mathrm{est}^{\vb{J}}}
\newcommand{\todo}[1]{ {\textcolor{red}{#1}} }
\newcommand{\R}[1]{R_{#1}}
\newcommand{\NN}[1]{\mathcal{N}\left(#1\right)}

\preprint{APS/123-QED}
\title{Supplemental Material for "Thermodynamic Uncertainty Relation in
  Interacting Many-Body Systems"\\~\\}

\author{Timur Koyuk and Udo Seifert\\~
}
\affiliation{
{II.} Institut f\"ur Theoretische Physik, Universit\"at Stuttgart,
  70550 Stuttgart, Germany
}
\date{\today}

\begin{abstract}
This supplemental material contains three sections. In the first section, we
define the currents entering the TUR. In the second
section, we explicitly calculate the inverses of the correlation matrices to
obtain the optimal estimate of entropy production using the MTUR. In the third section, we calculate the quality factors of the total particle
current and the total power for a mixture of two
species.
\end{abstract}
\maketitle
\section{I. Definition of Currents}
In this section, we will define the currents entering the TUR and the MTUR in
Eqs.~(1) and~(2) in the main text, respectively. Since these two relations are
valid for discrete and continuous systems, we will define currents for both
system types. For the sake of simplicity, we consider two-dimensional systems.
We first define
currents for a general two-dimensional overdamped Langevin equation and then
for the driven lattice gas as an example for a system with a discrete set of
states.

We consider $\N$ particles in two dimensions at positions
$\vb{r}\equiv\left(\vb{r}_1,...,\vb{r}_\N\right)$ with coordinates $\vb{r}_i\equiv (x_i,y_i)$ obeying the
overdamped Langevin equation
\begin{equation}
  \label{eq:suppl:Langevin_Eq}
  \partial_t\vb{r}_t\equiv \vb{\dot{r}}_t = \mat{\mu}\left[-\nabla V_\mathrm{int}(\vb{r}_t) + \vb{f}\right] + \sqrt{2}\mat{G}\vb{\zeta}_t.
\end{equation}
Here, $\mat{\mu}$ is the $2\N\times2\N$ mobility matrix, $\nabla V_\mathrm{int}(\vb{r}_t)$ is
the gradient of the interaction potential, $\vb{f}\equiv(\Xindex{\vb{f}}{1},...,\Xindex{\vb{f}}{\N})$ is a vector containing $\N$
non-conservative forces
$\Xindex{\vb{f}}{i}\equiv(\Xindex{f_x}{i},\Xindex{f_y}{i})$ with spatial
components $\Xindex{f_x}{i}$ and $\Xindex{f_y}{i}$, $\mat{G}$ is a $2\N\times2\N$
matrix used to define the symmetric diffusion matrix
$\mat{D}\equiv\mat{G}\mat{G}^\mathrm{T}=\mat{\mu}/\beta$ and
$\vb{\zeta}_t\equiv(\Xindex{\vb{\zeta}}{1}_t,...,\Xindex{\vb{\zeta}}{\N}_t)$ is a vector of $\N$
white Gaussian noises $\Xindex{\vb{\zeta}}{i}_t\equiv[\Xindex{\zeta_x}{i}(t), \Xindex{\zeta_y}{i}(t)]$ describing the random forces with mean and correlations
\begin{align}
  \label{eq:suppl:noise_mean}
  \expval{\Xindex{\zeta_x}{i}(t)} &= 0,\\
  \expval{\Xindex{\zeta_a}{i}(t)\Xindex{\zeta_b}{j}(t')} &= \delta_{a,b}\delta_{ij}\delta(t-t'),
  \label{eq:suppl:noise_correlation}  
\end{align}
respectively, where $a,b\in\{x,y\}$.
A general fluctuating current along the trajectory $\vb{r}_t$ of length $\T$ reads
\begin{equation}
  \label{eq:suppl:Langevin_current}
  J[\vb{r}_t] \equiv \frac{1}{\T}\int_0^{\T}\dd{t}\vb{d}(\vb{r}_t)\circ\vb{\dot{r}}_t,
\end{equation}
where
$\vb{d}(\vb{r}_t)\equiv[\Xindex{\vb{d}}{1}(\vb{r}_t),...,\Xindex{\vb{d}}{\N}(\vb{r}_t)]$
is a vector of arbitrary increments
$\Xindex{\vb{d}}{i}(\vb{r}_t)\equiv[\Xindex{d_x}{i}(\vb{r}_t),\Xindex{d_y}{i}(\vb{r}_t)]$
and $\circ$ denotes the Stratonovich
product. The choice $\vb{d}(\vb{r})= \beta\vb{f}$ in
Eq.~\eqref{eq:suppl:Langevin_current} corresponds to the total power
\begin{equation}
  \label{eq:suppl:Langevin_entropy_production} P[\vb{r}_t] \equiv
\frac{1}{\T}\int_0^{\T}\dd{t}\beta\vb{f}\circ\vb{\dot{r}}_t.
\end{equation}
Choosing $\vb{d}(\vb{r})=\Xindex{\vb{e}_x}{i}$ as the unit vector of particle
$i$ in direction $x$, we
get the current in x-direction of particle $i$ as
\begin{equation}
  \label{eq:suppl:Langevin_particle current}
  \Xindex{J}{i}[\vb{r}_t] \equiv
  \frac{1}{\T}\int_0^{\T}\dd{t}\dot{x}_i(t)
\end{equation}
with mean value $\Xindex{J}{i}\equiv\expval{\Xindex{J}{i}[\vb{r}_t]}$.

Next, we consider currents for the driven lattice gas discussed in the main
text. The fluctuating current of particle $i$ along the trajectory $\conf_t$
of length $\T$ reads
\begin{equation}
  \label{eq:suppl:DLG_particle_current}
  \Xindex{J}{i}\left[\conf_t\right] \equiv \frac{1}{\T}\left[\Xindex{n}{i}_{x^+}(\T)-\Xindex{n}{i}_{x^-}(\T)\right],
\end{equation}
where $\Xindex{n}{i}_{x^+}(\T)$ and $\Xindex{n}{i}_{x^-}(\T)$  denote the total number of jumps of particle
$i$ in positive and in negative $x$-direction up to time $\T$,
respectively. The mean value in Eq.~\eqref{eq:suppl:DLG_particle_current} is
defined as
$\Xindex{J}{i}\equiv\expval{\Xindex{J}{i}\left[\conf_t\right]}$.  Using
Eq.~\eqref{eq:suppl:DLG_particle_current}, we define the particle currents
of species $1$ and $2$ as
\begin{align}
  \label{eq:suppl:DLG_current_one}
  \Xtype{J}{1}\left[\conf_t\right]\equiv \frac{1}{\Xtype{N}{1}}\sum_{i=1}^\N \delta_{1,\alpha_i}
  \Xindex{J}{i}\left[\conf_t\right]
\end{align}
and
\begin{align}
  \Xtype{J}{2}\left[\conf_t\right]\equiv \frac{1}{\Xtype{N}{2}}\sum_{i=1}^\N \delta_{2,\alpha_i}
  \Xindex{J}{i}\left[\conf_t\right],
  \label{eq:suppl:DLG_current_two}  
\end{align}
respectively. We denote the corresponding mean values by
$\Xtype{J}{1}\equiv\expval{\Xtype{J}{1}\left[\conf_t\right]}$ and
$\Xtype{J}{2}\equiv\expval{\Xtype{J}{2}\left[\conf_t\right]}$. The total
power is given by
\begin{equation}
  \label{eq:suppl:DLG_entropy_production} P\left[\conf_t\right] \equiv
\beta \Xtype{q}{1} E \Xtype{\N}{1}\Xtype{J}{1}\left[\conf_t\right] + \beta \Xtype{q}{2} E
\Xtype{\N}{2}\Xtype{J}{2}\left[\conf_t\right]
\end{equation}
with mean value $P\equiv\expval{P[\conf_t]}=\beta \Xtype{q}{1} E\Xtype{\N}{1} \Xtype{J}{1}+ \beta \Xtype{q}{2} E
\Xtype{\N}{2}\Xtype{J}{2}$.
The total particle current reads
\begin{equation}
  \label{eq:suppl:DLG_total_particle_current}
  J_\mathrm{tot}\left[\conf_t\right] \equiv \sum_{i=1}^{\N} \Xindex{J}{i}\left[\conf_t\right]
\end{equation}
with mean value
$J_\mathrm{tot}\equiv\expval{J_\mathrm{tot}\left[\conf_t\right]}=\N_1\Xtype{J}{1}
+\N_2\Xtype{J}{2}$.

For currents in both, continuous and discrete systems, the diffusion coefficient and
the correlations between two particle currents are defined as
\begin{align}
  \label{eq:suppl:Langevin_diffusion_coefficient}
  \Xindex{D}{i} &= \T\mathrm{Var}[\Xindex{J}{i}]/2 \equiv
  \T\expval{(\Xindex{J}{i}[X_t]
                  -\Xindex{J}{i})^2}/2
\end{align}
and 
\begin{align}
  \Xindex{C}{ij} &= \T\mathrm{Cov}[\Xindex{J}{i},\Xindex{J}{j}]/2 \nonumber\\
                &\equiv \T
                   (\expval{\Xindex{J}{i}[X_t]\Xindex{J}{j}[X_t]} -
                  \Xindex{J}{i}\Xindex{J}{j})/2,
                    \label{eq:suppl:Langevin_correlations}
\end{align}
respectively, with $X_t\in\{\vb{r}_t,\conf_t\}$. The mean values of the power
in Eqs.~\eqref{eq:suppl:Langevin_entropy_production}
and~\eqref{eq:suppl:DLG_entropy_production} coincide with the mean total
entropy production, i.e., $\sigma\equiv\expval{P[X_t]}$. However, for any
finite time $\T$, their fluctuating values and consequently their diffusion
coefficients are different. In contrast, for long observation times
$\T\to\infty$, the total fluctuating power becomes the total entropy
production, i.e., $P\left[X_t\right]\approx\sigma\left[X_t\right]$, as the
contribution of the change in internal energy and in stochastic entropy
vanishes asymptotically.

\section{II. Inverse of the Correlation Matrix}
We explicitly calculate the inverse of the $\N\times\N$ correlation
matrix
\begin{equation}
  \label{eq:suppl:correlation_matrix} \C_{ij} \equiv \Xindex{D}{i} \delta_{i,j}
+ \Xindex{C}{ij}(1-\delta_{i,j})
\end{equation}
for a homogeneous system with one and two species to get the optimal
estimate for entropy production given by
\begin{equation}
  \label{eq:suppl:opt_estimate} \sigestMTUR \equiv \vb{J}^{T}\Cmat^{-1}\vb{J}.
\end{equation}

We note that the optimal estimate via the MTUR~\eqref{eq:suppl:opt_estimate} can be
derived by building the linear combination of the particle currents
$\mathcal{J}\equiv\sum_{i=1}^{\N}\varphi_i\Xindex{J}{i}$, inserting
$\mathcal{J}$ as a current into the ordinary TUR, Eq.~(1) in the main text,
and optimizing the estimate $\sigest^{\mathcal{J}}$ with respect to the
coefficients $\varphi_i\in\mathbb{R}$. The optimal estimator
$\sigest^{\mathcal{J^*}}$ of the optimized linear combination
$\mathcal{J^*}\equiv\sum_{i=1}^{\N}\varphi^*_i\Xindex{J}{i}$ with
coefficients $\varphi^*_i$ then coincides with Eq.~\eqref{eq:suppl:opt_estimate}.

\subsection{Homogeneous System with one Species}
As outlined in the main text, the mean
values $\Xindex{J}{i} \equiv J$, diffusion coefficients $\Xindex{D}{i} \equiv
D$ and correlations $\Xindex{C}{ij} \equiv C$ are independent of $i$ and $j$ for a homogeneous
system. Thus, the correlation matrix~\eqref{eq:suppl:correlation_matrix} reads
\begin{equation}
  \label{eq:suppl:hom_sys_correlation_matrix}
  \C_{ij} \equiv D \delta_{i,j} +
C(1-\delta_{i,j}).
\end{equation}
The inverse of this matrix is given by
\begin{equation}
  \label{eq:suppl:inv_hom_sys_correlation_matrix}
  \C^{-1}_{ij} = \frac{(D/C + N - 2)\delta_{ij} - (1-\delta_{ij})
}{ [D/C-1]\left[(D-C) +\N C\right]},
\end{equation}
which can be easily verified by calculating $\sum_j
\C^{-1}_{ij}\C_{jk}=\delta_{ik}$. Inserting
Eq.~\eqref{eq:suppl:inv_hom_sys_correlation_matrix}
into~\eqref{eq:suppl:opt_estimate} yields the optimal estimate of entropy
production, i.e., Eq.~(4) in the main text. Since this estimate is identical
to the estimate obtained when choosing the entropy production as a current in
the ordinary TUR, the optimal coefficients $\varphi^*_i$ are all identical,
i.e., $\varphi^*_i = \varphi^*_j$ for all $i,j$.

\subsection{Homogeneous System with two Species}
Since the system is homogeneous, the currents $\Xindex{J}{i}=
\Xtype{J}{\alpha_i}$, the diffusion coefficients
$\Xindex{D}{i}=\Xtype{D}{\alpha_i}$ and the correlations
$\Xindex{C}{ij}=\Xtype{C}{\alpha_i \alpha_j}$ are identical within a
species. Therefore, we split the vector $\vb{J}=(\Xtype{\vb{J}}{1},
\Xtype{\vb{J}}{2} )$ into the vectors $\Xtype{\vb{J}}{1}$ and $\Xtype{\vb{J}}{2}$, which
contain $\Xtype{N}{1}$ entries of $\Xtype{J}{1}$ and $\Xtype{N}{2}$ entries of
$\Xtype{J}{2}$, respectively. Moreover, we can split the correlation matrix
into submatrices, i.e.,
\begin{equation}
  \label{eq:suppl:two_species_sub_mat}
  \Cmat = 
  \begin{pmatrix}
    \Xtype{\Cmat}{11} & \Xtype{\Cmat}{12} \\
\Xtype{\Cmat}{12}^\mathrm{T}& \Xtype{\Cmat}{22}
\end{pmatrix}
\end{equation}
with the $\Xtype{\N}{1}\times\Xtype{\N}{1}$ matrix $\Xtype{\Cmat}{11}$, the
$\Xtype{\N}{2}\times\Xtype{\N}{2}$ matrix $\Xtype{\Cmat}{22}$ and the the
$\Xtype{\N}{1}\times\Xtype{\N}{2}$ matrix $\Xtype{\Cmat}{12}$. The entries of
the matrices $\Xtype{\Cmat}{11}$ and $\Xtype{\Cmat}{22}$ read
\begin{equation}
  \label{eq:suppl:C_11}
  \left(\Xtype{\Cmat}{11}\right)_{ij} = \Xtype{D}{1} \delta_{i,j} +
\Xtype{C}{11}(1-\delta_{i,j})
\end{equation}
and
\begin{equation}
  \label{eq:suppl:C_22}
  \left(\Xtype{\Cmat}{22}\right)_{ij} = \Xtype{D}{2} \delta_{i,j} +
\Xtype{C}{22}(1-\delta_{i,j}),
\end{equation}
respectively. Both matrices consist of correlation functions between particles within
a species, whereas the matrix $\Xtype{\Cmat}{12}$ with entries
\begin{equation}
  \label{eq:suppl:C_12}
  \left(\Xtype{\Cmat}{12}\right)_{ij} = \Xtype{C}{12} \delta_{i,j} +
\Xtype{C}{12}(1-\delta_{i,j})
\end{equation}
consists of correlation functions between particles of different species.

To proof Eq.~(6) in the main text, we present two alternatives. We first show
heuristically that the problem of inverting the $\N\times\N$--matrix in
Eq.~\eqref{eq:suppl:two_species_sub_mat} can be simplified to inverting a
$2\times2$ matrix. Then, we show a more rigorous way to invert the
matrix~\eqref{eq:suppl:two_species_sub_mat} by using the block-matrix
inversion formula.

As we have previously derived for a homogeneous system with one
species, the optimal coefficients $\varphi^*_j$ are all identical within a
species.  As a consequence, each current within a species contributes with the
same weight to the optimal estimate for entropy production. Thus, the optimal
estimate reads
\begin{equation}
  \label{eq:suppl:two_species_2x2_optimal_estimate} \sigestMTUR =
\vb{J'}^\mathrm{T} \Cmat'^{-1} \vb{J'}
\end{equation} with $\vb{J'}\equiv\left(\Xtype{J}{1},\Xtype{J}{2}\right)$ and
the $2\times2$--matrix
\begin{equation}
  \label{eq:suppl:two_species_matrix_C} \Cmat' =
  \begin{pmatrix} \Xtype{\N}{1}\Xtype{\eta}{1} & \Xtype{\N}{1} \Xtype{N}{2} \Xtype{C}{12}\\
\Xtype{\N}{1} \Xtype{N}{2} \Xtype{C}{12} & \Xtype{\N}{2}\Xtype{\eta}{2}
  \end{pmatrix},
\end{equation} where $\Xtype{\eta}{\alpha} \equiv \Xtype{D}{\alpha}
+ (\N_\alpha-1)\Xtype{C}{\alpha\alpha}$ and
$\alpha\in\set{1,2}$. Calculating the inverse of the matrix in
Eq.~\eqref{eq:suppl:two_species_matrix_C} and inserting it
into~\eqref{eq:suppl:two_species_2x2_optimal_estimate} leads to Eq.~(6) in the
main text.

For a more rigorous proof of the optimal estimate in Eq.~(6) in the main text,
we have to explicitly calculate the inverse of the $\N\times\N$--matrix in
Eq.~\eqref{eq:suppl:two_species_sub_mat} by using the block-matrix inversion
formula. For a $\N\times\N$--matrix
\begin{equation}
  \label{eq:suppl:matrix_definition_block_inversion}
  \mat{M} \equiv
  \begin{pmatrix} \mat{A} & \mat{B} \\
    \mat{C}& \mat{D}
   \end{pmatrix}  
 \end{equation}
 the block-inversion formula reads
 \begin{widetext}
\begin{equation}
  \label{eq:suppl:block_matrix_inversion}
  \mat{M}^{-1} =
  \begin{pmatrix} \mat{A}^{-1} + \mat{A}^{-1}\mat{B}\left(\mat{D}-\mat{C}\mat{A}^{-1}\mat{B}\right)^{-1}\mat{C}\mat{A}^{-1} & -\mat{A}^{-1}\mat{B}\left(\mat{D}-\mat{C}\mat{A}^{-1}\mat{B}\right)^{-1} \\
    -\left(\mat{D}-\mat{C}\mat{A}^{-1}\mat{B}\right)^{-1}\mat{C}\mat{A}^{-1} & \left(\mat{D}-\mat{C}\mat{A}^{-1}\mat{B}\right)^{-1}
   \end{pmatrix}.
 \end{equation}
 \end{widetext}
Using Eq.~\eqref{eq:suppl:block_matrix_inversion} to calculate the inverse of
the matrix in~\eqref{eq:suppl:two_species_sub_mat} and inserting it
into~\eqref{eq:suppl:opt_estimate} yields
\begin{align}
  \label{eq:suppl:est_two_species_intermediate}
  \sigestMTUR =
  &\Xtype{\vb{J}}{1}^\mathrm{T}\Xtype{\Cmat}{11}^{-1}\Xtype{\vb{J}}{1} +
  \Xtype{\vb{J}}{1}^\mathrm{T}\Xtype{\Cmat}{11}^{-1}\Xtype{\Cmat}{12} \mat{K}^{-1}\Xtype{\Cmat}{12}^\mathrm{T}\Xtype{\Cmat}{11}^{-1}\Xtype{\vb{J}}{1}\nonumber\\
  &-\Xtype{\vb{J}}{1}^\mathrm{T}\Xtype{\Cmat}{11}^{-1}\Xtype{\Cmat}{12}\mat{K}^{-1}\Xtype{\vb{J}}{2} \nonumber\\
    &-\Xtype{\vb{J}}{2}^\mathrm{T}\mat{K}^{-1}\Xtype{\Cmat}{12}^\mathrm{T}\Xtype{\Cmat}{11}^{-1}\Xtype{\vb{J}}{1}\nonumber\\
  &+ \Xtype{\vb{J}}{2}^\mathrm{T}\mat{K}^{-1}\Xtype{\vb{J}}{2}
\end{align}
with
\begin{equation}
  \label{eq:suppl:K_matrix}
  \mat{K}\equiv \Xtype{\Cmat}{22}
-\Xtype{\Cmat}{12}^\mathrm{T}\Xtype{\Cmat}{11}^{-1}\Xtype{\Cmat}{12}.
\end{equation}
Analogously to Eq.~\eqref{eq:suppl:inv_hom_sys_correlation_matrix}, the inverse of the matrix $\Xtype{\Cmat}{11}$ is given by
\begin{equation}
  \label{eq:suppl:inverse_C_11}
  (\Xtype{\Cmat}{11}^{-1})_{ij} = \frac{(\Xtype{D}{1}/\Xtype{C}{11} + \Xtype{N}{1} - 2)\delta_{ij} - (1-\delta_{ij})
  }{[\Xtype{D}{1}/\Xtype{C}{11}-1]\left[(\Xtype{D}{1}-\Xtype{C}{11}) +\Xtype{\N}{1}\Xtype{C}{11}\right]}.
\end{equation}
and the inverse of matrix $\mat{K}$ is given by
 \begin{widetext}
\begin{equation}
  \label{eq:suppl:K_matrix}
  K^{-1}_{ij} = \frac{[(\Xtype{D}{2}-\Xtype{m}{12})/(\Xtype{C}{22}-\Xtype{m}{12}) + \Xtype{\N}{2} - 2]\delta_{ij} - (1-\delta_{ij})
}{[(\Xtype{D}{2}-\Xtype{m}{12})/(\Xtype{C}{22}-\Xtype{m}{12})-1]\left[\{(\Xtype{D}{2}-\Xtype{m}{12})-(\Xtype{C}{22}-\Xtype{m}{12})\} + \Xtype{N}{2} (\Xtype{C}{22}-\Xtype{m}{12})\right]}
\end{equation}
\end{widetext}
with
\begin{equation}
  \label{eq:suppl:m_12}
\Xtype{m}{12}\equiv
\frac{ \Xtype{C}{12}^2 \Xtype{\N}{1} } { \Xtype{D}{1}-\Xtype{C}{11} +
  \Xtype{N}{1}\Xtype{C}{11} }.
\end{equation}
Inserting the inverses~\eqref{eq:suppl:inverse_C_11}
and~\eqref{eq:suppl:K_matrix} into
Eq.~\eqref{eq:suppl:est_two_species_intermediate}, evaluating all terms and
writing them into a symmetric form with respect to the particle species yields
finally Eq.~(6) in the main text.

\section{III. Quality Factors}
In this section, we derive the quality factor of the total power
$\Q_P$ and of the total particle current $\Q_{J_\mathrm{tot}}$ for a
mixture of two particle species and show that, in general, these quality factors
differ from the quality factor of the MTUR $\QMTUR$.

For a mixture of two particle species, the total entropy production reads
\begin{equation}
  \label{eq:suppl:Q_def_sigma}
  \sigma \equiv \beta \Xtype{f}{1} \Xtype{N}{1} \Xtype{J}{1} + \beta \Xtype{f}{2} \Xtype{N}{2} \Xtype{J}{2},
\end{equation}
where $\Xtype{f}{1,2}$ are non-conservative forces acting on particle species
$1$ and $2$, respectively. To derive the quality factors, we use the identity
\begin{equation}
  \label{eq:suppl:Q_fluctuation_identity}
  \mathrm{Var}\left[\mathcal{J}\right] = \sum_{i=1}^{\N}
  \varphi^2_i\mathrm{Var}\left[\Xindex{J}{i}\right] + 2\sum_{i>j} \varphi_i\varphi_j\mathrm{Cov}\left[\Xindex{J}{i},\Xindex{J}{j}\right]
\end{equation}
for a current
$\mathcal{J}\left[\conf_t\right]\equiv\sum_{i=1}^\N\varphi_i\Xindex{J}{i}\left[\conf_T\right]$. Using~\eqref{eq:suppl:Q_fluctuation_identity}
for evaluating the diffusion coefficient $D_{\mathcal{J}}$ and choosing
$\varphi_i = \Phi_{\alpha_i}$ as arbitrary species dependent increments
$\Phi_{\alpha_i}\in\{\Xtype{\Phi}{1},\Xtype{\Phi}{2}\}$ yields for homogeneous systems
\begin{align}
  D_{\mathcal{J}} =& \Xtype{\Phi}{1}^2\Xtype{\N}{1}\Xtype{D}{1} +
                     \Xtype{\Phi}{2}^2\Xtype{\N}{2}\Xtype{D}{2}\nonumber\\
  &+\Xtype{\Phi}{1}^2\Xtype{N}{1}(\Xtype{N}{1}-1)\Xtype{C}{11}
  +\Xtype{\Phi}{2}^2\Xtype{N}{2}(\Xtype{N}{2}-1)\Xtype{C}{22}\nonumber\\
  &+ 2\Xtype{\Phi}{1}\Xtype{\Phi}{2}\Xtype{\N}{1}\Xtype{\N}{2}\Xtype{C}{12}.
    \label{eq:suppl:Q_diffusion_coefficient_D_J}
\end{align}
Choosing $\Xtype{\Phi}{1}=\Xtype{f}{1}$ and $\Xtype{\Phi}{2}=\Xtype{f}{2}$
leads to the estimator of the total power
\begin{widetext}
\begin{equation}
  \label{eq:suppl:sigest_sigma}
  \sigest^P \equiv P^2/D_P = \frac{\left(\beta \Xtype{f}{1} \Xtype{N}{1}
    \Xtype{J}{1} + \beta \Xtype{f}{2} \Xtype{N}{2} \Xtype{J}{2}\right)^2}{
    \beta^2\Xtype{f}{1}^2\Xtype{\N}{1}\Xtype{D}{1} + \beta^2\Xtype{f}{2}^2\Xtype{\N}{2}\Xtype{D}{2}
    +\beta^2\Xtype{f}{1}^2\Xtype{N}{1}(\Xtype{N}{1}-1)\Xtype{C}{11}
    +\beta^2\Xtype{f}{2}^2\Xtype{N}{2}(\Xtype{N}{2}-1)\Xtype{C}{22}
    + 2\beta^2\Xtype{f}{1}\Xtype{f}{2}\Xtype{\N}{1}\Xtype{\N}{2}\Xtype{C}{12}}
\end{equation}
and choosing $\Xtype{\Phi}{1}=\Xtype{\Phi}{2}=1$ leads to the estimator
of the total particle current
\begin{equation}
  \label{eq:suppl:sigest_J_tot}
  \sigest^{J_\mathrm{tot}}\equiv J^2_\mathrm{tot}/D_{J_\mathrm{tot}} = \frac{\left(\Xtype{N}{1}
    \Xtype{J}{1} + \Xtype{N}{2} \Xtype{J}{2}\right)^2}{
    \Xtype{\N}{1}\Xtype{D}{1} + \Xtype{\N}{2}\Xtype{D}{2}
    + \Xtype{N}{1}(\Xtype{N}{1}-1)\Xtype{C}{11}
    + \Xtype{N}{2}(\Xtype{N}{2}-1)\Xtype{C}{22}
    + 2\Xtype{\N}{1}\Xtype{\N}{2}\Xtype{C}{12}}.
\end{equation}
\end{widetext}
By using the estimators Eqs.~\eqref{eq:suppl:sigest_sigma}
and~\eqref{eq:suppl:sigest_J_tot} and taking the thermodynamic limit
$\N\to\infty$, we get the the quality factor of the total power
\begin{widetext}
\begin{equation}
  \label{eq:suppl:Q_sigma}
  \Q_P \equiv \sigest^P/\sigma = \frac{\beta \Xtype{f}{1} \Xtype{\rho}{1}
    \Xtype{J}{1} + \beta \Xtype{f}{2} \Xtype{\rho}{2} \Xtype{J}{2}}{
    \beta^2\Xtype{f}{1}^2\Xtype{\rho}{1}\Xtype{D}{1} + \beta^2\Xtype{f}{2}^2\Xtype{\rho}{2}\Xtype{D}{2}
    +\beta^2\Xtype{f}{1}^2\Xtype{\rho}{1}\Xtype{\gamma}{11}
    +\beta^2\Xtype{f}{2}^2\Xtype{\rho}{2}\Xtype{\gamma}{22}
    + 2\beta^2\Xtype{f}{1}\Xtype{f}{2}\Xtype{\rho}{1}\Xtype{\rho}{2}\Xtype{\gamma}{12}}
\end{equation}
and the quality factor of the total particle current
\begin{equation}
  \label{eq:suppl:Q_J_tot}
  \Q_{J_\mathrm{tot}} \equiv \sigest^{J_\mathrm{tot}}/\sigma = \frac{\left(\Xtype{\rho}{1}
    \Xtype{J}{1} + \Xtype{\rho}{2} \Xtype{J}{2}\right)^2}{
    \left(\Xtype{\rho}{1}\Xtype{D}{1} + \Xtype{\rho}{2}\Xtype{D}{2}
    +\Xtype{\rho}{1}\Xtype{\gamma}{11}
    +\Xtype{\rho}{2}\Xtype{\gamma}{22}
    + \Xtype{\rho}{1}\Xtype{\rho}{2}\Xtype{\gamma}{12}\right)\left(\beta \Xtype{f}{1} \Xtype{\rho}{1}
    \Xtype{J}{1} + \beta \Xtype{f}{2} \Xtype{\rho}{2} \Xtype{J}{2}\right)}
\end{equation}
\end{widetext}
in the thermodynamic limit. The quality factors~\eqref{eq:suppl:Q_sigma}
and~\eqref{eq:suppl:Q_J_tot} are in general different than the quality factor
based on the MTUR in Eq.~(7) in the main text. However, in the one-species
case $\Xtype{\rho}{2}=0$, these quality factors coincide with the quality factor
based on the MTUR [cf. Eq.~(5) in the main text].